\documentclass{elsart}

\usepackage{amssymb}

\begin{document}

\begin{frontmatter}

\title{Shadows of a Maximal Acceleration}

\author{G. Papini}

\address{Department of Physics, University of Regina, Regina, Sask. S4S 0A2, Canada}
\address{International Institute for Advanced Scientific Studies, 84019 Vietri sul Mare
(SA), Italy}

\begin{abstract}
A quantum mechanical upper limit on the value of particle accelerations, or maximal
acceleration (MA), is applied to compact stars. A few MA fermions are at most present in
canonical white dwarfs and neutron stars. They drastically alter a star's stability
conditions.
\end{abstract}

\end{frontmatter}

  Caianiello argued in 1984 that
Heisenberg's uncertainty relations place an upper limit  ${\mathcal A}_{m}$ on the value
that the acceleration of a particle can take along a worldline \cite{cai2}.  This limit,
referred to as maximal acceleration (MA), is determined by the particle's mass itself. It
is distinct from the value that has been derived in some works \cite{br,cai3,gasp} from
the Planck mass $ m_{P}= \left(\hbar c/G \right) ^{\frac{1}{2}}$ and that is therefore a
universal constant. With some modifications \cite{wood} and additions, Caianiello's
argument goes as follows.

If two observables $ \hat{f}$ and $ \hat{g}$ obey the commutation relation
\begin{equation}\label{al}
\left[\hat{f},\hat{g}\right]= - i \hbar \hat{\alpha},
\end{equation}
where $ \hat{\alpha}$ is a Hermitian operator, then their uncertainties
\begin{eqnarray}
 \left(\Delta f\right)^{2}&=&
<\Phi\mid\left(\hat{f}-<\hat{f}>\right)^{2}\mid\Phi>\\ \nonumber
 \left(\Delta g\right)^{2}&=&
<\Phi\mid\left(\hat{g}-<\hat{g}>\right)^{2}\mid\Phi>
\end{eqnarray}
 also satisfy the inequality
\begin{equation}
\left(\Delta f\right)^{2}\cdot\left(\Delta g\right)^{2}\geq
\frac{\hbar^{2}}{4}<\Phi\mid\hat{\alpha}\mid\Phi>^{2},
\end{equation}
or
\begin{equation}\label{be}
\Delta f\cdot\Delta g\geq \frac{\hbar}{2} \mid<\Phi\mid\hat{\alpha}\mid\Phi>\mid .
\end{equation}
Using Dirac's analogy between the classical Poisson bracket $ \left\{f,g\right\}$ and the
quantum commutator \cite{land}
\begin{equation}
\left\{f,g\right\} \rightarrow\frac{1}{i\hbar}\left[\hat{f},\hat{g}\right],
\end{equation}
one can take $ \hat{\alpha}=\left\{f,g\right\}\hat{\mathbf{1}}$. With this substitution,
Eq.(\ref{al}) yields the usual momentum-position commutation relations. If in particular
$ \hat{f}=\hat{H}$, then Eq.(\ref{al}) becomes
\begin{equation}
\left[\hat{H},\hat{g}\right]= - i \hbar \left\{H,g\right\}\hat{\mathbf{1}},
\end{equation}
(\ref{be}) gives \cite{land}
\begin{equation}\label{&}
\Delta E\cdot \Delta g \geq \frac{\hbar}{2} \mid \left\{H,g\right\}\mid
\end{equation}
and
\begin{equation}\label{&&}
\Delta E\cdot\Delta g\geq \frac{\hbar}{2}\mid\frac{dg}{dt}\mid ,
\end{equation}
when $ \frac{\partial g}{\partial t}=0 $. Eqs.(\ref{&}) and (\ref{&&}) are re-statements
of Ehrenfest theorem. Criteria for its validity are discussed at length in the literature
\cite{mess,land,bal}. Eq.(\ref{&&}) implies that $\Delta E =0$ when the quantum state of
the system is an eigenstate of $\hat{H}$. In this case $\frac{dg}{dt}= 0$.

If $g \equiv v(t)$ is the (differentiable) velocity expectation value of a particle whose
energy is $ E$, then Eq.(\ref{&&}) gives
\begin{equation}\label{7}
\mid \frac{dv}{dt}\mid \leq \frac{2}{\hbar} \Delta E \cdot \Delta v(t) .
\end{equation}
 In general \cite{sha}
\begin{equation}
 \Delta v =
\left(<v^{2}>-<v>^{2}\right)^{\frac{1}{2}}\leq v_{max}\leq c .
\end{equation}
 Caianiello's additional assumption, $ \Delta E \leq E$, remains unjustified.
He used this assumption to estimate $ E$ by observing that the acceleration is largest in
the {\it instantaneous} rest frame of a particle where $ E = mc^{2}$.  If, in fact, the
three-acceleration in the instantaneous rest frame is $ \vec{a}_{p}$, then the particle's
acceleration $ \vec{a}' $ in another frame with instantaneous velocity $ \vec{v}$ is
given by \cite{steph}
\begin{equation}\label{quater}
\vec{a}'=
\frac{1}{\gamma^{2}}\left[\vec{a}_{p}-\frac{\left(1-\gamma\right)\left(\vec{v}\cdot
\vec{a}_{p}\right)\vec{v}}{v^{2}}-\frac{\gamma \left(\vec{v}\cdot
\vec{a}_{p}\right)\vec{v}}{c^{2}}\right] .
\end{equation}
 The equation
\begin{equation}\label{quinquies}
a'^{2}= \frac{1}{\gamma^{4}}\left(a_{p}^{2}-\frac{\left(\vec{a}_{p}\cdot
\vec{v}\right)^{2}}{c^{2}}\right),
\end{equation}
where $ a^{2} \equiv \vec{a}\cdot \vec{a}$, follows from (\ref{quater}) and shows that $
a' \leq a_{p}$ for all $ \vec{v}\neq 0$ and that $ a' \rightarrow 0 $ as $ \mid
\vec{a}_{p}\cdot \vec{v}\mid \rightarrow a_{p}c $. Caianiello then concluded that in a
frame of reference in which the particle is "nearly at rest"
\begin{equation}\label{8}
\mid \frac{dv}{dt}\mid \leq 2\frac{mc^{3}}{\hbar}\equiv {\mathcal A}_{m} .
\end{equation}

 It also follows that in the instantaneous rest frame of
the particle the absolute value of the proper acceleration is
\begin{equation}\label{9}
\left(\mid
\frac{d^{2}x^{\mu}}{ds^{2}}\frac{d^{2}x_{\mu}}{ds^{2}}\mid\right)^{\frac{1}{2}}=
\left(\mid\frac{1}{c^{4}}\frac{d^{2}x^{i}}{dt^{2}}\mid \right)^{\frac{1}{2}}\leq
\frac{\mathcal{A}_{m}}{c^{2}} .
\end{equation}
Eq.(\ref{9}) is a Lorentz invariant. The validity of (\ref{9}) under Lorentz
transformations is therefore assured.

Caianiello's argument can also be extended to include the average length of the
acceleration $ <a>$. If, in fact, $ v(t)$ is differentiable, then fluctuations about its
mean are given by
\begin{equation}\label{del}
\Delta v \equiv v-<v>\simeq \left(\frac{dv}{dt}\right)_{0}\Delta
t+\left(\frac{d^{2}v}{dt^{2}}\right)_{0}\left(\Delta t\right)^{2}+...  .
\end{equation}
Eq.(\ref{del}) reduces to
\begin{equation}\label{sh}
\Delta v\simeq \mid\frac{dv}{dt}\mid \Delta t
=<a> \Delta t
\end{equation}
 for sufficiently small values of $\Delta t$, or when
$\mid\frac{dv}{dt}\mid$ remains constant over $\Delta t$. Heisenberg uncertainty relation
\begin{equation}\label{yy}
\Delta E \cdot \Delta t\geq \hbar/2 ,
\end{equation}
and (\ref{sh}) then yield
\begin{equation}\label{ga}
<a> \leq \frac{2mc^{3}}{\hbar}.
\end{equation}
 In the absence of counterexamples, results (\ref{8}) and (\ref{ga}) have almost
acquired the status of a principle.

The notion of MA delves into a number of issues and is connected to the extended nature
of particles. In fact, the inconsistency of the point particle concept for a relativistic
quantum particle is discussed by Hegerfeldt \cite{heg} who shows that the localization of
the particle at a given point at a given time conflicts with causality.

Classical and quantum arguments supporting the existence of MA have been frequently
discussed in the literature
\cite{mthw,das,gasp1,toller,paren,vora,mash,venzo,falla,pati}. MA would eliminate
divergence difficulties affecting the mathematical foundations of quantum field theory
\cite{nester}. It would also free black hole entropy of ultraviolet divergences
\cite{hoof,suss,mcgui}. MA plays a fundamental role in Caianiello's geometrical
formulations of quantum mechanics \cite{cai4} and in the context of Weyl space
\cite{pap}. A limit on the acceleration also occurs in string theory. Here the upper
limit appears in the guise of Jeans-like instabilities that develop when the acceleration
induced by the background gravitational field is larger than a critical value for which
the string extremities become casually disconnected \cite{sanchez,gasp2,gasp3}. Frolov
and Sanchez \cite{fro} have also found that a universal critical acceleration must be a
general property of strings.

An important question regarding MA is how to incorporate it in a theory that take into
account the limits (\ref{8}) or (\ref{9}) in a meaningful way from inception. The model
of Ref.\cite{cai4} represents a step in this direction and has led to interesting results
that range from particle physics to astrophysics and cosmology \cite{pap1,cai3,cai5}.
This question will not however be discussed here.

A second, equally fundamental question, is how MA makes itself manifest in Nature in ways
that are model independent. This is the objective of the present work.

The limits (\ref{8}) and (\ref{9}) are of course very high for most particles (for an
electron $ {\mathcal A}_{m}\sim 4.7\times 10^{31} cm s^{-2}$) and likely to occur only in
exceptional physical circumstances. An example is discussed below.

It regards the highly unusual conditions of matter in the interior of white dwarfs and
neutron stars. In this case the legitimacy of (\ref{&&}) is assured by the inequality $
\frac{N}{V} \lambda_{T}^{3}\ll 1$, where $ \frac{N}{V}$ is the particle density in the
star and $ \lambda_{T}= \frac{2\pi\hbar^{2}}{mkT}$ is of the same order of magnitude of
the de Broglie wavelength of a particle with a kinetic energy of the order of $ kT$
\cite{land,bal}.

Consider the simple model of a star consisting of $N$ free fermions constrained to move
in a spherical box. The model applies, with due changes, to both white dwarfs and neutron
stars. The details are in general given for the former case only. Because the electron
energy in the interior of the star is $E\sim 20 MeV$, the evolution time $\Delta t$ to
distinguishable states is sufficiently small to justify the application of (\ref{ga}).
The standard expression for the ground state energy of an ideal fermion gas inside a star
of radius $R$ and volume $V$ is
\begin{equation}\label{27}
E_{0}(r)= \frac{m^{4}c^{5}}{\pi^{2}\hbar^{3}}V f(x_{F}),
\end{equation}
where $ x_{F}\equiv \frac{p_{F}}{mc}$, $ p_{F}\equiv
\left(3\pi^{2}\frac{N}{V}\right)^{\frac{1}{3}}\hbar$ is the Fermi momentum, $ 0\leq r\leq
R$, $N$ is the total number of fermions and
\begin{equation}\label{28}
f( x_{F}) = \int_{0}^{x_{F}}dx x^{2}\sqrt{1 + x^{2}}.
\end{equation}
The integral in (\ref{28}) is approximated by
\begin{equation}\label{29}
f(x_{F})\simeq \frac{1}{3}x_{F}^{3}\left(1 + \frac{3}{10}x_{F}^{2}+...\right),  x_{F}\ll
1
\end{equation}
or by
\begin{equation}\label{30}
f(x_{F})\simeq \frac{1}{4}x_{F}^{4}\left(1 + \frac{1}{x_{F}}^{2}+...\right),  x_{F}\gg 1
.
\end{equation}
The two cases are usually referred to as non-relativistic (NR) and extreme relativistic
(ER) respectively. The average force exerted by the fermions a distance $r$ from the
center of the star is
\begin{equation}\label{31}
<F_{0}> = \frac{\partial E_{0}}{\partial r}= \frac{4m^{4}c^{5}}{\pi\hbar^{3}}r^{2}f(
x_{F})+\frac{m^{4}c^{5}}{\pi^{2}\hbar^{3}}V\frac{\partial f}{\partial
x_{F}}\frac{\partial x_{F}}{\partial r} .
\end{equation}
The last term in (\ref{31}) corresponds, in the case of white dwarfs, to electrostatic
interactions among ions and to inverse $\beta$-decay corrections. They are neglected here
in first approximation. The resulting equation of state is therefore polytropic
\cite{shap}. From (\ref{31}) and the expression
\begin{equation}\label{32}
N( r) = \frac{4}{9\pi\hbar^{3}}r^{3}{p_{F}}^{3}
\end{equation}
that gives the number of fermions in the ground state at $ r$, one finds the average
acceleration per fermion as a function of $ r$
\begin{equation}\label{33}
<a(r)> = \frac{9c^2}{{x_{F}}^{3}r}f(x_{F}).
\end{equation}
By using (\ref{29}) and (\ref{30}) one finds to second order
\begin{eqnarray}\label{34}
<a(r)>_{NR}&=& \frac{3c^{2}}{r}\left(1+\frac{3}{10}x_{F}^{2}\right)\\ \nonumber
<a(r)>_{ER}&=&\frac{9c^{2}}{4r}\left(x_{F}+\frac{1}{x_{F}}\right).
\end{eqnarray}
$<a(r)>$ must now satisfy the MA limit (\ref{ga}). One finds
\begin{eqnarray}\label{35}
r\geq (r_{0})_{NR}&\equiv& \frac{3\lambda}{4\pi}\\ \nonumber r\geq
(r_{0})_{ER}&\equiv&\frac{9}{16\pi}\frac{\lambda p_{F}}{mc} ,
\end{eqnarray}
where $\lambda\equiv\frac{h}{mc} $ is the Compton wavelength of $m$. For a white dwarf, $
N/V\simeq 4.6\times 10^{29}  cm^{-3}$ gives $ (r_{0})_{NR}\simeq 5.8\times 10^{-11} cm$
and $ (r_{0})_{ER}\simeq 4\times 10^{-11} cm$.

The corresponding values for a gas of neutrons are $ (r_{0})_{NR}\simeq 3.2\times
10^{-14} cm$ and $ (r_{0})_{ER}\simeq 2.2\times 10^{-14} cm$.

 In order to have at least one state with particles reaching MA values, one must
have
\begin{eqnarray}\label{36}
Q\left((r_{0})_{NR}\right)&=&\frac{4}{9\pi}\left((r_{0})_{NR}\frac{p_{F}}{\hbar}\right)^{3}\sim
1 \\ \nonumber
Q\left((r_{0})_{ER}\right)&=&\frac{4}{9\pi}\left((r_{0})_{ER}\frac{p_{F}}{\hbar}\right)^{3}\sim
1 .
\end{eqnarray}
In the case of a typical white dwarf, the first of (\ref{36}) gives $ (N/V)_{NR}\sim
1.2\times 10^{30} cm^{-3}$ and the second one $(N/V)_{ER}\sim 1.3\times 10^{30} cm^{-3}$.
On the other hand, the condition $x_{F}\ll 1$ requires $ (N/V)_{NR}\ll 6\times 10^{29}
cm^{-3}$, whereas $x_{F}\gg 1$ yields $(N/V)_{ER}\gg 6\times 10^{29} cm^{-3}$. It
therefore follows that the NR approximation does not lead to electron densities
sufficient to produce states with MA electrons. The possibility to have states with MA
electrons is not ruled out entirely in the ER case.

The corresponding densities for a typical neutron star are $(N/V)_{NR}\sim 3\times
10^{42} cm^{-3}$ and $(N/V)_{ER}\sim 3\times 10^{39} cm^{-3}$, whereas $x_{F}\ll 1$
requires $(N/V)_{NR}\ll \frac{8\pi}{3\lambda^{3}}\sim 3.6\times 10^{42} cm^{-3}$ and
$x_{F}\gg 1$ leads to $(N/V)_{ER}\gg 3.6 \times 10^{42} cm^{-3}$. In both instances,
states with MA neutrons do not seem possible.

The outlook is however different if one starts from conditions that do not lead
necessarily to the formation of "canonical" white dwarfs or neutron stars. Pressure is
the essential ingredient. It consists of two terms. The first term represents the
pressure exerted by the small fraction of particles that can attain accelerations
comparable with MA. It is given by
\begin{equation}\label{37}
P_{MA}= \frac{2m^{2}c^{3}}{\hbar}\frac{Q(r_{0})}{4\pi r_{0}^{2}} .
\end{equation}
The second part is the contribution of those fermions in the gas ground state that can
not achieve MA
\begin{equation}\label{38}
-\frac{\partial E_{0}^{\prime}}{\partial V}= -\frac{\partial}{\partial
V}\left(2\tilde{\gamma} V \int_{r_{0}}^{\infty}d\epsilon
\frac{\epsilon^{3/2}}{e^{\alpha+\beta}+1}\right) ,
\end{equation}
where $ \tilde{\gamma} \equiv
\frac{\left(2m\right)^{3/2}}{\left(2\pi\right)^{2}\hbar^{3}}$ and $
V=\frac{4}{3}\pi\left(R^{3}-r_{0}^{3}\right)$. Since $r_{0}$ is small, one can write
\begin{equation}\label{39}
\frac{\partial E_{0}^{\prime}}{\partial V}\simeq \frac{\partial E_{0}}{\partial V}\sim
\frac{4K\tilde{M}^{5/3}}{5\tilde{R}^{5}} ,
\end{equation}
where $ \tilde{M}\equiv \frac{9\pi M}{8m_{p}}$, $ \tilde{R}\equiv \frac{2\pi
R}{\lambda}$, $K\equiv \frac{m^{4}c^{5}}{12\pi^{2}\hbar^{3}}$ and $m_{p}$ is the mass of
the proton. In the non relativistic case, the total pressure is obtained by adding
(\ref{37}) to (\ref{39}) and using the appropriate expression for $Q(r_{0})$ in
(\ref{36}). The hydrostatic equilibrium condition is
\begin{equation}\label{40}
\frac{8\pi m
c^{2}}{3\lambda^{3}}\frac{\tilde{M}}{\tilde{R}^{3}}+\frac{4K}{5}\frac{\tilde{M}^{5/3}}
{\tilde{R}^{5}}= K^{\prime}\frac{\tilde{M}^{2}}{\tilde{R}^{4}},
\end{equation}
where $K^{\prime}\equiv \frac{4\alpha G \pi}{\lambda^{4}}\frac{64 m_{p}^{2}}{81}$ and
$\alpha \approx 1$ is a factor that reflects the details of the model used to describe
the hydrostatic equilibrium of the star. It is generally assumed that the configurations
taken by the star are polytropes \cite{shap}. Solving (\ref{40}) with respect to $
\tilde{R}$ one finds
\begin{equation}\label{41}
\tilde{R}=\frac{\tilde{M} \tilde{M_{0}}^{-2/3}}{8}\left(1 \mp \sqrt{1-\frac{64}{5}
\left(\frac{\tilde{M}_{0}}{\tilde{M}}\right)^{4/3}} \right) ,
\end{equation}
where $ \tilde{M}_{0}\equiv \left(K/K^{\prime}\right)^{3/2}= \left(\frac{27\pi\hbar
c}{64\alpha G m_{p}}\right)^{\frac{3}{2}}\simeq \frac{9\pi}{8m_{p}}M_{\odot}$. Solutions
(\ref{41}) will be designated by $ \tilde{R}_{-}$ and $ \tilde{R}_{+}$. They are real if
$ M\geq \left(\frac{64}{5}\right)^{3/4}M_{0}\sim 6.8 M_{\odot}$. The corresponding
electron densities, calculated from $ N/V = \frac{3M}{8\pi m_{p}R^{3}}$, are
$\left(N/V\right)_{-}
> \frac{8\pi}{3}\left(\frac{8}{\lambda}\right)^{3}\left(M_{0}/M\right)^{2}\sim 6.6\times
10^{30}cm^{-3}$ and $\left(N/V\right)_{+}< 6.6\times 10^{30}cm^{-3}$. The electron
density for $R_{+}$ is still compatible with that of a canonical NR white dwarf.
Eq.(\ref{40}) can also be written in the form
\begin{equation}\label{42}
\tilde{M}^{1/3} \tilde{R}= \frac{4}{5}\tilde{M}_{0}^{2/3}\left(1+\frac{10\pi m c^{2}}{3
\lambda^{3}K}\frac{\tilde{R}^{2}}{\tilde{M}^{2/3}}\right) ,
\end{equation}
where the second term on the r.h.s. represents the MA contribution to the usual M-R
relation for NR white dwarfs. This contribution can be neglected when $
\frac{\tilde{R}}{\tilde{M}^{1/3}}< 1/\sqrt{5}$ which requires $ N/V > \frac{8\pi
5^{3/2}}{3 \lambda^{3}} \sim 6.6 \times 10^{29} cm^{-3}$. This condition and the usual
M-R relation, $ \tilde{M}^{1/3}\tilde{R}=\frac{4\tilde{M}_{0}^{2/3}}{5} $, are compatible
if $ R < \frac{\lambda}{\pi 5^{3/4}}\left(\frac{9\pi M_{0}}{8m_{p}}\right)^{1/3}$, which
leads to the density $N/V
>\frac{5^{9/4}\pi}{3\lambda^{3}}\left(M/M_{0}\right)\sim 2.7\times 10^{30}(M/M_{0})
cm^{-3}$.

Similarly, one can calculate the MA contribution to the M-R relation for ER white dwarfs.
The equation is
\begin{equation}\label{43}
\frac{m^{4}c^{5}}{4 \pi^{2}\hbar^{3}}+K
\left(\frac{\tilde{M}^{4/3}}{\tilde{R}^{4}}-\frac{\tilde{M}^{2/3}}{\tilde{R}^{2}}\right)
=K^{\prime}\frac{\tilde{M}^{2}}{\tilde{R}^{4}} ,
\end{equation}
where the first term on the l.h.s. represents the MA contribution. From (\ref{43}) one
gets
\begin{equation}\label{44}
\tilde{R}=\tilde{M}^{1/3}\sqrt{4-\left(\frac{\tilde{M}}{\tilde{M}_{0}}\right)^{2/3}} ,
\end{equation}
and the stability condition becomes $ M\leq 8 M_{0}\sim 8 M_{\odot}$. For the electron
densities determined, the star can still be called a white dwarf. One also finds that for
$N/V\sim 3\times 10^{30}cm^{-3}$ the number of MA states is only $
Q(r_{0})=\frac{9\lambda^{3}}{16\pi^{2}}\frac{N}{V}\simeq 2.2\frac{M}{M_{\odot}}$. A few
MA electrons could therefore be present at this density. However, interactions involving
electrons and protons at short distances may occur before even this small number of
electrons reaches the MA.

Analogous conclusions also apply to neutron stars, with minor changes if these can be
treated as Newtonian polytropes. This approximation may only be permissible, however, for
low-density stars \cite{shap}. One finds in particular $ Q(r_{0})=\frac{9
\lambda_{n}^{3}}{16\pi^{2}}\frac{N}{V}\simeq 4.5 \frac{M}{M_{\odot}}$. The presence of a
few MA neutrons is therefore allowed in this case.

In most instances the value of MA is so high to defy direct observation and become almost
irrelevant. Nonetheless the role of MA as a possible sort of universal regulator must not
be discounted. It is an intrinsic, first quantization limit that preserves the continuity
of spacetime and does not require the introduction of a fundamental length, or of
arbitrary cutoffs. The challenge is to find physical situations where MA is relatively
lower and its presence somewhat perceived.

The physical situation considered regards matter in the interior of white dwarfs and
neutron stars. For canonical white dwarfs, the possibility that states exist with MA
electrons can be ruled out in the NR case, but not so for ER stars. For canonical neutron
stars the possibility is ruled out in both NR and ER instances. On the other hand, the
mere presence of a few MA electrons alters the stability conditions of the white dwarf
drastically, as shown by (\ref{42}) and (\ref{44}). This also applies to NR neutron
stars, with some provisos, however, regarding the choice of the equation of state. In the
collapse of stars with masses larger than the Chandrasekhar and Oppenheimer-Volkoff
limits from white dwarfs to neutron stars, to more compact objects, conditions favorable
to the formation of states with MA fermions may occur for the time allowed by competing
processes like phase changes.

This research was supported by the Natural Sciences and Engineering Research Council of
Canada.

\end{document}